\documentclass[11pt,leqno]{article}
\usepackage{amsmath,amssymb,amsbsy,amsfonts,amsthm,mathtools,mathrsfs}
\usepackage[T1]{fontenc} 			
\usepackage{times}				
\usepackage[normalem]{ulem}			
\usepackage{scrextend}				
\usepackage{calc}					
\usepackage{comment}				
\usepackage{enumitem}				
\usepackage{scalerel}				
\usepackage{fontawesome}			
\usepackage{mathabx}				
\usepackage{bbm}					
\usepackage{authblk}				
\usepackage[letterpaper, top=1in, bottom=1in, left=.75in, right=.75in]{geometry}
\usepackage{setspace}				
\usepackage[pdfencoding=auto,psdextra,colorlinks=TRUE,linkcolor=black,citecolor=black,urlcolor=blue]{hyperref}
\usepackage{bookmark}				

\newfont{\boldit}{cmbxti12}
\newcommand{\R}{\mathbbm{R}}  						
\newcommand{\deq}{\:\dot{=}\:}						
\newcommand{\lb}{\langle}							
\newcommand{\rb}{\rangle}							
\newcommand{\LB}{\left\langle}						
\newcommand{\RB}{\right\rangle}						

\newcommand{\CkdC}[2]{C^{#1}_{0}(\R^{#2})}				
\newcommand{\DTo}{\mathcal{D}(\R)}						
\newcommand{\DTd}[1]{\mathcal{D}(\R^{#1})}				
\newcommand{\DDd}[1]{\mathcal{D}\,'(\R^{#1})}				
\newcommand{\Lpd}[2]{L^{#1}(\R^{#2})}					
\newcommand{\Locd}[2]{L_{\text{loc}}^{#1}(\R^{#2})}			
\newcommand{\fall}{\;\;\forall\;}						
\newcommand{\deltap}{\delta\,'}						

\renewcommand{\geq}{\vargeq}

\renewcommand{\thefootnote}{\fnsymbol{footnote}}

\makeatletter
\def\thmheadbrackets#1#2#3{%
  \thmname{#1}\thmnumber{\@ifnotempty{#1}{ }\@upn{#2}}%
  \thmnote{ {\the\thm@notefont[\;#3\;]}}}
\makeatother

\newtheoremstyle{brakets}
  {1.5em}
  {1.5em}
  {\itshape}
  {}
  {\bfseries}
  {\\}
  { }
  {\thmheadbrackets{#1}{#2}{#3}}

\theoremstyle{brakets}
\newtheorem{thm}{Theorem}[section]
\newtheorem{cor}[thm]{Corollary}

\newtheorem{lem}[thm]{Lemma}
\newtheorem{prop}[thm]{Proposition}

\numberwithin{equation}{section}

\title{The Dirac Delta as a Singular Potential for the 2D Schrodinger Equation}
\author{Michael Maroun\footnote{Chief Scientist and VP of R\&D, TeXDyn Industries Corporate Laboratories}}
\affil{TeXDyn Industries Corporate Laboratories \\
Austin, TX \slash\; Boston, MA}
\date{\today}

\begin{document}
\maketitle

\renewcommand{\thefootnote}{\arabic{footnote}}

\begin{abstract}
In the framework of distributionally generalized quantum theory, the object $H\psi$ is defined as a distribution. The mathematical significance is a mild generalization for the theory of para- and pseudo-differential operators (as well as a generalization of the weak eigenvalue problem), where the $\psi$-do symbol (which is not a proper linear operator in this generalized case) can have its coefficient functions take on singular distributional values. Here, a distribution is said to be singular if it is not $\Lpd{p}{d}$ for any $p\geq 1$. Physically, the significance is a mathematically rigorous method, which does not rely upon renormalization or regularization of any kind, while producing bound state energy results in agreement with the literature. In addition, another benefit is that the method does not rely upon self-adjoint extensions of the Laplace operator. This is important when the theory is applied to non-Schrodinger systems, as is the case for the Dirac equation and a necessary property of any finite rigorous version of quantum field theory. The distributional interpretation resolves the need to evaluate a wave function at a point where it fails to be defined. For $d=2$, this occurs as $K_o(a|x|)\delta(x)$, where $K_o$ is the zeroth order MacDonald function. Finally, there is also the identification of a missing anomalous length scale, owing to the scale invariance of the formal symbol(ic) Hamiltonian, as well as the common identity for the logarithmic function, with $a,\,b\in\R^+$, $\log(ab)=\log(a)+\log(b)$, which loses unitlessness in its arguments. Consequently, the energy or point spectrum is generalized as a family (set indexed by the continuum) of would-be spectral values, called the C-spectrum.
\end{abstract}

\section{Introduction}
The central abstract symbol of study, which is not a linear operator but for which one would like some notion of bound state and analogue of point spectrum (i.e. energy eigenvalue(s)), is
\[
H = -\frac{\hbar^2}{2m}\sum\limits_{j=1}^d\frac{\partial^2}{\partial x_j^2} - \alpha\delta(x),
\]
where $m,\, \hbar > 0$, $\alpha\in\R\setminus\{0\}$, and $x\in\R^d$. The mathematical hurdle that is overcome is the fact that a priori, the object $H$ does not define a linear operator, since the potential is not a function and hence cannot be a bona fide operator of multiplication on the usual infinite dimensional square integrable Hilbert space defined through Lebesgue measure. In the present work here, the dimension $d$ is set equal to 2. Some results will be specific to $d=2$, while others will be more general. Comments are made accordingly for each. Careful distinction is made between the Laplace operator and the distributional Laplace operator, which have very different properties. The usual Laplace operator will be denoted $\nabla^2 = \frac{\partial^2}{\partial x^2} + \frac{\partial^2}{\partial y^2}$, while the distributional Laplace operator will be denoted $\Delta\deq\partial_x^2 + \partial_y^2$. The binary relation $\dot{=}$ will mean equal in the sense of distributions, throughout. The distinction between the two types of Laplace operators is vital becomes the usual Laplace operator is an unbounded linear operator, whereas the distributional Laplace operator is not only bounded (and hence continuous) as a linear operator, but it also is necessarily not a monotone operator. In physical language, the distributional Laplace operator undergoes rapid oscillations\footnote{The property of rapid oscillations was pointed out by M. L. Lapidus in a personal communication.}.

\section{A Symmetry of the Singular Hamiltonian}

It is useful to begin with some important properties of the symmetry of $H$. It is immediate that the object given by
\[
H = -\frac{\hbar^2}{2m}\Delta - \alpha\delta(x),
\]
is reflection invariant. Informally, one may write unambiguously $H(x) = H(-x)$ for $x\in\R^2$, since the second order nature of the Laplace operator (both distributional and usual) and the even property of the Dirac delta assures this relation. Less obviously, there is the scale invariance of $H$.
\begin{prop}[ Scale Invariance of the Symbolic Hamiltonian ] \label{SIoH}
Let $x\in\R^2$, $s\in\R\setminus\{0\}$, and $\phi\in\DTd{2}$ then,
\[
\langle H(sx),\,\phi\rangle = \langle H(x),\,\phi\rangle
\]
\end{prop}
\begin{proof}[ Proof of the Scale Invariance of H ]
Let $x\in\R^2$ and $s\in\R\setminus\{0\}$. For a vector $x\in\R^2$ the substitution $x = sx'$ for $x'\in\R^2$ represents an isotropic rescaling of the two dimensional Euclidean space. Let $\phi\in\DTd{2}$ be an infinitely smooth test function with compact support. The consequence on the Hamiltonian is then,
\begin{align*}
\langle H(x),\,\phi\rangle &= \LB-\frac{\hbar^2}{2m}\Delta(x) - \alpha\delta(x),\,\phi\RB \\
&= \LB-\frac{\hbar^2}{2m}\Delta(sx') - \alpha\delta(sx'),\,\phi\RB \\
&= \LB-\frac{\hbar^2}{2m}\frac{1}{s^2}\Delta(x') - \alpha\frac{1}{s^2}\delta(x'),\,\phi(x)\RB \\
&= \int\limits_{x\in\R^2}\frac{1}{s^2}\left[-\frac{\hbar^2}{2m}\Delta(x') - \alpha\delta(x')\right]\phi\;d^2x \\
&= \int\limits_{x'\in\R^2}\frac{1}{s^2}\left[-\frac{\hbar^2}{2m}\Delta(x') - \alpha\delta(x')\right]\phi s^2 d^2x' \\
&= \int\limits_{x'\in\R^2}\left[-\frac{\hbar^2}{2m}\Delta(x') - \alpha\delta(x')\right]\phi\;d^2x' \\
&= \langle H(sx'),\,\phi\rangle = \langle H(x'),\,\phi\rangle
\end{align*}
Hence $\langle H(sx),\,\phi\rangle = \langle H(x),\,\phi\rangle$ as claimed above. \qedhere
\end{proof}

\section{Some Special Results from the Theory of Distributions}

The following gives an inconclusive result, and is metaphorically similar to a failure of linear independence. For example, consider the trivial algebraic relation $x = x + 1$. The false statement that results from this tells one that the two graphs never intersect, and indeed after more geometric considerations one realizes this is due to the lines being parallel, while not colinear.
\begin{align*}
\langle\log\!|x|\,\delta(x),\,\phi\rangle &= -\langle\log\!|x|\,x\deltap(x),\,\phi\rangle \\
&= \langle\delta(x),\,\frac{d}{dx}[x\log\!|x|\,\phi]\rangle \\
&= \langle\delta(x),\,[\log\!|x|\,\phi + \phi + x\log\!|x|\,\phi\,']\rangle \\
&= \langle\log\!|x|\,\delta(x),\,\phi\rangle + \langle\delta(x),\,\phi(x)\rangle + \langle\delta(x),\,x\log\!|x|\,\phi\,'(x)\rangle \\
&= \langle\log\!|x|\,\delta(x),\,\phi\rangle + \langle\delta(x),\,\phi(x)\rangle + 0
\end{align*}
Since the same term appears on both the left and right hand side, the last line implies $\langle\delta(x),\,\phi\rangle = 0$ for all $\phi\in\DTo$, which of course is false. A similar contradiction arises in the exact same calculation for $d>1$ in $\R^d$. To understand more precisely what is happening, consider the following more generic calculation.
\begin{align*}
\langle f(x)\delta(x),\,\phi\rangle &= -\langle f(x) x \deltap(x),\,\phi\rangle \\
&= \langle\delta(x),\,\frac{d}{dx}[x f(x) \phi]\rangle \\
&= \langle\delta(x),\,[f(x)\phi + xf'(x)\phi + xf(x)\phi\,']\rangle
\end{align*}
The last line again has the first term on the right cancel with the original term on the left. Consequently, one is left with
\[
\langle xf'(x)\delta(x),\,\phi\rangle = \langle(f(x)\delta + xf'(x)\delta + xf(x)\deltap),\,\phi\rangle
\]
Again like terms on the left and right cancel and one is left with, $\langle(f(x)\delta + xf(x)\deltap),\,\phi\rangle = 0$, which gives the well known distributional identity for $x\in\R$, namely $\delta\;\dot{=} -x \deltap$, but in the form of a strong equality.
\[
\langle f(x)\delta,\,\phi\rangle = \langle -xf(x)\deltap,\,\phi\rangle
\]
It is now apparent that different choices of $f(x)$, especially interesting and possibly singular expressions such as $f(x)=\tfrac{1}{x}$, lead to novel statements. But the novel statements hinge upon a generic property of $f$ that it and its derivatives are related by finite sums, products and powers of $f$ itself. For the case of the logarithm, $f(x)=\log\!|x|\,$, this is not true as the logarithm cannot be expressed as finite sums, products and powers of its derivatives.

In the non-Banach infinite dimensional vector space of distributions, $\DDd{d}$, the binary operation of products does not exist, in general. This means that multiplication by the zero vector is not the usual ring annihilator, since there is no binary operation to make this have meaning. A logical and geometric consequence is that the zero distribution is simultaneously both {\it parallel} and {\it perpendicular} to itself. Owing to the non-Banach nature of $\DDd{d}$, there is no norm, and even with semi-norms (since $\DDd{d}$ is Fr\'{e}ch\'{e}t), the zero vector is not normalizable. So, the zero distribution is neither positive nor negative; it is simultaneously and uniquely the only symmetric {\it and} antisymmetric linear functional; finally, it is both perpendicular and parallel to itself.

It may be possible to obtain consistent novel results by considering the primatives of $\log\!|x|\,=x\log\!|x|\,-x$ and so on; but such calculations become quickly cumbersome and circular. Therefore, a different method of proof is desirable. Throughout, the notation $\delta(x-a)$ and $\delta_a$ are used interchangeably to mean pure point Dirac measure with singleton set of support at the point $x=a$.
\begin{lem}[ Triviality and Existence of $\delta_o\log$ ] \label{TnE}
For $x\in\R^d$, and $\phi\in\DTd{d}=\CkdC{\infty}{d}$, the expression $\log\!|x|\,\delta(x)$ is a distribution $\forall\phi\in\DTd{d}$. Specifically, it is the zero distribution, i.e.
\[
\lb\log\!|x|\,\delta_o,\,\phi\rb = 0,\; \fall\phi\in\DTd{d}
\]
\end{lem}
\begin{proof}[ Proof v1 of Lemma \ref{TnE} ]
By direct calculation, one obtains,
\begin{align*}
\langle\log\!|x|\,\delta(x),\,\phi(x)\rangle &= \int\limits_{x\in\R^d}\log\!|x|\,\delta(x)\phi(x) d^dx \\
&= C\int\limits_{0}^\infty\log(r)\frac{\delta(r)}{r^{d-1}}\varphi(r) r^{d-1} dr \\
&= C\int\limits_{0}^\infty\log(r)\delta(r)\varphi(r) dr \\
&= C\int\limits_{-\infty}^\infty t\,\delta(e^t)\, e^t\, \varphi(e^t)\; dt \qquad let\quad r = e^t\\
&= C\lim_{t\to-\infty} t\,e^t\,\varphi(e^t) \\
&= C\cdot 0\cdot\varphi(0) = 0 \qquad\fall\phi\in\DTd{d}. \text{\qedhere}
\end{align*}
\end{proof}
\noindent In the above, $C$ is an arbitrary constant, which depends upon only $\phi$. For example, if $\phi$ is rotationally invariant then $C=1$ owing to the result of the $d-1$ angular integrals canceling the surface area $\Omega_{d-1}$ of the d-dimensional unit ball, which appears in the denominator of the Dirac delta measure (not shown above) when expressed in d-spherical coordinates.

It turns out that there is another proof by direct calculation that obtains the same result in a different manner, but is specific to $d=2$. Since the quantum system under investigation is for $x\in\R^2$, this alternative proof is relevant, and is provided below.
\begin{proof}[ Proof v2 of Lemma \ref{TnE} for $d=2$ ]
The trick is to use the fundamental solution of the Laplace equation in $\R^2$, namely the distributional identity, $\Delta\log\!|x|\,\deq 2\pi\delta_o$.
\begin{align*}
\langle\log\!|x|\,\delta(x),\,\phi(x)\rangle &= \frac{1}{2\pi}\langle\log\!|x|\,\Delta\log\!|x|\,,\,\phi(x)\rangle \\
&= \frac{1}{2\pi}\langle\log\!|x|\,,\,\nabla^2[\log\!|x|\,\phi(x)]\rangle \\
&= \frac{1}{2\pi}\LB\log\!|x|\,,\,\left[0 + \frac{2}{|x|^2}x\cdot\nabla\phi(x) + \log\!|x|\,\nabla^2\phi(x)\right]\RB \\
&= \frac{1}{2\pi}\LB 0 - \frac{2}{|x^2|} + \Delta(\log\!|x|\,)^2,\, \phi(x)\RB \\
&= \frac{1}{2\pi}\LB-\frac{2}{|x^2|} + \frac{2}{|x|^2} + 4\pi\log\!|x|\,\delta(x),\, \phi(x)\RB \\
&= \langle 2\log\!|x|\,\delta(x),\, \phi(x)\rangle
\end{align*}
\noindent Consequently, $\langle\log\!|x|\,\delta(x),\,\phi(x)\rangle = \langle 2\log\!|x|\,\delta(x),\, \phi(x)\rangle$ implies $\langle\log\!|x|\,\delta(x),\, \phi(x)\rangle = 0$, as was to be shown. \qedhere
\end{proof}
The consequences of the trivial existence of $\delta_o\log$ are vital to the solution of the 2 dimensional Schrodinger system with singular Dirac potential. Specifically, the solution of the system entails attributing meaning to the singular distribution $K_o\delta_o$, where $K_o$ is the MacDonald function of zeroth order. This is the content of the following central theorem.
\begin{thm}[ $K_o\delta_o$ as a Distribution ] \label{KoDaD}
Let $x\in\R^2$, $|\cdot|$ the 2d Euclidean norm, $\phi\in\DTd{d}$, $L,\,a>0$, $\gamma$ the Euler-Mascheroni constant, and $K_o$ the zeroth order MacDonald function, then
\[
\lb K_o(a|x|)\delta(x),\,\phi\rb = \LB\,-\log\left(\frac{1}{2}\,e^\gamma\,a\,L\right)\delta(x),\,\phi\RB.
\]
\end{thm}
\noindent Aside from Lemma \ref{TnE}, there are 2 more facts needed to prove this result. The first is the asymptotic expansion at the origin of the zeroth order MacDonald function. It is given by,
\begin{equation}
K_o(a|x|)\overset{x\to0}{\sim} -\log\left(\frac{1}{2}\,e^\gamma\,a\,|x|\right) + \mathcal{O}(x^2). \label{eq1}
\end{equation}
The second fact seems like a pedantic nuisance but in fact is critical to recognizing the existence of the missing (hence anomalous) length scale in this quantum system. Note that self-adjoint extensions of the Laplace operator do not expose this fact. The anomaly appears as a consequence of the identity $\log(ab) = \log(a) + \log(b)$ for $a,\,b\in\R^+=(0,\infty)$. One can see if the product $(ab)$ is unitless, then the right hand side is the sum of mathematical functions whose arguments are not unitless, which is a problem. This is one of the many hidden sources of renormalization in quantum physics and quantum field theory. One very simple resolution to this unavoidable fact is to introduce a positive constant $L>0$ with units such that whenever the product $ab$ is unitless, the respective products $aL$ and $bL^{-1}$ are also unitless. Ergo, one obtains the modified statement,
\[
\log\left(a\,\frac{L}{L}\,b\right) = \log(aL) + \log\left(\frac{b}{L}\right),
\]
with the somewhat strange caveat that one must not again apply the mathematical identity lest the reapplication create a sequence of arbitrary positive constants $L_j>0$ for all $j\in\{1,2,3,...,n\}$, where $n$ is the number of times the identity gets recursively applied. A one parameter family indexing the anomaly suffices, on account of the symmetry of scale invariance demonstrated in proposition \ref{SIoH}.
\begin{proof}[ Proof of Theorem \ref{KoDaD} ]
Throughout this proof, one has $x\in\R^2$, $|\cdot|$ the 2d Euclidean norm, and $\phi\in\DTd{2}$. From the asymptotic form at the origin, the MacDonald function can be replaced by its asymptotic relation \ref{eq1},
\begin{align*}
\lb K_o(a|x|)\delta(x),\,\phi\rb &= \LB -\log\left(\tfrac{1}{2}\,e^\gamma\,a\,|x|\right)\delta(x),\,\phi\RB \\
&= \LB-\log\left(\tfrac{1}{2}\,e^\gamma\,a\tfrac{L}{L}\,|x|\right)\delta(x),\,\phi\RB \\
&= \LB\left[-\log\left(\tfrac{1}{2}\,e^\gamma\,a\,L\right) - \log\left(\tfrac{|x|}{L}\right)\right]\delta(x),\,\phi\RB \\
&= \LB-\log\left(\tfrac{1}{2}\,e^\gamma\,a\,L\right)\delta(x),\,\phi\RB - \int\limits_{x\in\R^2}\log\left(\tfrac{|x|}{L}\right)\delta(x)\,\phi(x)d^2x \qquad\text{let}\;x=Lx' \\
&= \LB-\log\left(\tfrac{1}{2}\,e^\gamma\,a\,L\right)\delta(x),\,\phi\RB - \int\limits_{x'\in\R^2}\log\left(|x'|\right)\delta(x')\,\phi(Lx')d^2x' \\
&= \LB-\log\left(\tfrac{1}{2}\,e^\gamma\,a\,L\right)\delta(x),\,\phi\RB - \lb\,\log|x'|\,\delta(x'),\,\phi_L(x')\,\rb \\
\end{align*}
\noindent In the last equality, the second term on the right is zero as a cause of lemma \ref{TnE}. Therefore, one is left with,
\[
\lb K_o(a|x|)\delta(x),\,\phi\rb = \lb-\log\left(\tfrac{1}{2}\,e^\gamma\,a\,L\right)\delta(x),\,\phi\rb,
\]
which proves theorem \ref{KoDaD}. \qedhere
\end{proof}

\section{Recapitulation of Generalized Quantum Theory}

In \cite{Mar1,Mar2} and \cite{Mar3}, the author creates framework for a mathematically rigorous generalization of quantum theory, which is based in the theory of distributions. The framework introduces the notion of a generalized spectrum, called the C-spectrum and reformulates the stationary quantum eigenvalue problem for the Schrodinger equation so that it can accommodate singular distributions. Here, a distribution is said to be singular if it is not $\Lpd{p}{d}$ for some $p\geq 1$. This is not drastically different than the standard definition. In the standard literature, a distribution is singular if it is not regular. Then, a distribution, $T_f$, is regular if $\exists f\in\Locd{1}{d}$ such that $T_f(\phi) = \lb f,\,\phi\rb < \infty\fall\phi\in\DDd{d}$.

In the Schrodinger Hamiltonian, the kinetic term of the Hamiltonian has the Laplace operator replaced with the distributional Laplace operator, which is a continuous linear operator unlike the usual Laplace operator. It then becomes natural to embed the formal symbol $(H\psi)$ inside the infinite dimensional topological vector space of distributions given by the continuous topological dual to the vector space of infinitely differentiable functions with compact support on $\R^d$ denoted, $\DTd{d}=\CkdC{\infty}{d}$. Thus, one writes the dual space of distributions as $\DDd{d}$. Since the test function space is contained in the space of square integrable Lebesgue functions characterizing the usual infinite dimensional Hilbert space of standard quantum theory, the Gel'fand triple with strict containment is central, $\CkdC{\infty}{d}=\DTd{d}\subset\Lpd{2}{d}\subset\DDd{d}$.

While the notion of the Gel'fand triple dates back to the 1960s where it was used extensively by authors A. Bohm and M. Gadella and by I. Gel'fand, G. Shilov, and N. Vilenkin whose first author's name it bears, the indispensable nature of the concept earned it the name `rigged Hilbert space'. The crux of generalized quantum theory is whether the formal symbol $(H\psi)$ is an element of the space of distributions. This fact depends heavily upon whether the potential denoted $V(x)$ defines a product, with the wave function $\psi(x)$ that is also itself a distribution. Note that it is not guaranteed a priori that $(V\psi)$ is a distribution because in general products of distributions do not define another distribution. In the present work, the central theorem \ref{KoDaD} is vital, as it establishes that
\[
(V\psi)(x) \deq \delta_o\,K_o \deq -\log\!|ax|\,\delta(x)\,\psi(x) \deq (-\log(a)\,\delta_o + 0)\in\DDd{2},
\]
for some $a>0$, and hence $(H\psi)\in\DDd{2}$. To wit, this brings up the present quantum system under consideration.

The formal symbol $(H\psi)$ is verified to be a rigorous element of the space of distributions as follows. One solves the Schrodinger equation with no potential but for the fundamental solution, $G$, such that $-\Delta G(x) + b^2 G(x) \deq \delta(x)$, with $G\in\Lpd{2}{2}$, and $b^2=\tfrac{2mE}{\hbar^2}$, when $E<0$. For $x\in\R^2$, $G(x) = \tfrac{1}{2\pi}K_o(b|x|)$. This gives the normalized wave function as,
\begin{equation}\label{psi}
\psi(x) = \frac{b}{\sqrt{\pi}}K_o(b|x|)\in\Lpd{2}{2},
\end{equation}
with $b^2 = \frac{2m|E|}{\hbar^2}>0$. Then application of the distributional Laplace operator produces,
\begin{equation}
\Delta\psi(x) \deq b^2\psi(x) - 2\pi\delta(x).
\end{equation}
One now has the second central theorem.
\begin{thm}[ $(H\psi)\in\DDd{2}$ ]
For $\psi\in\Lpd{2}{2}$ as in \ref{psi}, $\hbar,\,m>0$, and $\alpha\in\R\setminus\{0\}$, one has $(H\psi)\in\DDd{2}$.
\end{thm}
\begin{proof}
\begin{align*}
H\psi(x) &= -\frac{\hbar^2}{2m}\Delta\psi(x) - \alpha\delta(x)\psi(x) \\
&= -\frac{\hbar^2}{2m}b^2\psi(x) + \frac{\hbar^2b}{m\sqrt{\pi}}\delta(x) - \alpha\delta(x)\psi(x) \\
&= E\,\psi(x) + \frac{\hbar^2 b}{m\sqrt{\pi}}\delta(x) - \frac{\alpha b}{\sqrt{\pi}}\delta(x)K_o(b|x|)
\end{align*}
For the last line, one must use the results of theorem \ref{KoDaD} to replace $\delta(x)K_o(b|x|)$ with $-\log\left(\tfrac{1}{2}e^\gamma bL\right)\delta(x)$, which gives
\begin{equation}\label{Hpsi}
H\psi(x) \,\dot{=} E\,\psi(x) + \frac{\hbar^2b}{m\sqrt{\pi}}\delta(x) + \frac{\alpha b}{\sqrt{\pi}}\log\left(\tfrac{1}{2}e^\gamma bL\right)\delta(x)
\end{equation}
Since $\psi\in\Lpd{2}{2}$ and $\delta\in\DDd{2}$, so is its sums, and thus $(H\psi)\in\DDd{2}$. \qedhere
\end{proof}
In the framework of generalized quantum theory the stationary states are found by equating $\lb H\psi,\,\phi\rb = E\lb\psi,\,\phi\rb$. Equation \ref{Hpsi} gives $\lb H\psi,\,\phi\rb = \LB \left[E\,\psi(x) + \frac{\hbar^2b}{m\sqrt{\pi}}\delta(x) + \frac{\alpha b}{\sqrt{\pi}}\log\left(\tfrac{1}{2}e^\gamma bL\right)\delta(x)\right],\,\phi\RB = E\lb\psi,\,\phi\rb$. From which, one must conclude,
\[
\LB\left[\frac{\hbar^2\pi}{m} + \alpha\log\left(\tfrac{1}{2}e^\gamma bL\right)\right]\!\frac{b}{\sqrt{\pi}}\,\delta_o,\,\phi\RB = 0
\]
Since in general $\lb\delta_o,\,\phi\rb = \phi(0)\neq 0$, this leads to
\begin{equation}\label{Eeq}
\frac{\hbar^2\pi}{m} + \alpha\log\left(\tfrac{1}{2}e^\gamma bL\right) = 0.
\end{equation}
Recalling $E = -\frac{\hbar^2}{2m}b^2 < 0$ for bound states, and solving for the energy through the parameter $b$ in the above equation \ref{Eeq}, yields,
\begin{equation}
E = -\frac{2\hbar^2}{mL^2}e^{-2\gamma - \tfrac{2\pi\hbar^2}{m\alpha}},
\end{equation}
where $m,\,\hbar,\,L>0$, $\alpha\in\R\setminus\{0\}$, and $\gamma = -\Psi(1)$ is the Euler-Mascheroni constant with $\Psi$ the digamma function with its usual definition as $\Gamma(x)\Psi(x) = \frac{d}{dx}\Gamma(x)$, where $\Gamma(x)$ is the gamma function defined by its functional equation $z\Gamma(z) = \Gamma(z+1)$. Thus, there is the C-spectrum, which is a one parameter family parametrized by the anomalous length scale originally such that $L>0$, but now extended so that,
\[
\sigma_c(H\psi) = \left\{-\frac{2\hbar^2}{mL^2}e^{-2\gamma - \tfrac{2\pi\hbar^2}{m\alpha}}\right\}_{L\in\R\setminus\{0\}}.
\]
\noindent Due to this extension, one should regard $\log(\cdot)$ as $\log|\cdot|$, where necessary.

The final result is to show that the solution here matches exactly that in the mathematics literature found in the definitive reference on point interactions for the Schrodinger theory, \cite{AGHH}. One has,
\begin{cor}[ Accordance with Self-Adjoint Extensions ]
When $\tfrac{\hbar^2}{m} = 2$ and $L=\pm 1$, the C-spectral family becomes the singleton set,
\[
\sigma^o_C(H\psi) = \left\{-4e^{-2\gamma - \tfrac{4\pi}{\alpha}}\right\},
\]
and the point spectrum $\sigma_p$ on page 100 in theorem 5.4, equation 5.25 of \cite{AGHH} coincides precisely with the above singleton set, i.e. $\sigma_p = \sigma^o_C$, provided $\alpha$ here is replaced with $\tfrac{1}{\text{\boldmath$\alpha$}}$, where $\text{\boldmath$\alpha$}$ is the alpha in \cite{AGHH}.
\end{cor}
The proof is straightforward with the trickiest part being the conclusion that the coupling constant in \cite{AGHH} is upside down from the one defined in the traditional Schrodinger system as a multiplicative constant in front of the Dirac delta as a singular potential. That this is true is found by computing the 2d scattering length. But the result is briefly quoted on the top of page 99's second sentence in \cite{AGHH}, namely $(-2\pi\text{\boldmath$\alpha$})^{-1}$ represents the scattering length. For this to be true, the alpha ($\alpha$) in this work must have units equal to the multiplicative inverse of the alpha ($\text{\boldmath$\alpha$}$) in \cite{AGHH}.

\section*{Acknowledgments}
The author is indebted to Distinguished Professor of Mathematics Michel L. Lapidus for his helpful comments and persistent welcoming of the author to his research seminar on Fractal Dynamics and Mathematical Physics at the University of California, Riverside.

\end{document}